\documentclass[%
 reprint,
 amsmath,amssymb,
 aps,
]{revtex4-1}

\usepackage{graphicx}
\usepackage{dcolumn}
\usepackage{bm}

\begin{document}
\preprint{APS/123-QED}
\title{Stochastic switching in Rydberg atomic ensemble}

\author{Jun He$^{1,2}$}
\author{Xin Wang$^{1}$} 
\author {Xin Wen$^{1}$}
\author{Junmin Wang$^{1,2,}$}
 \email{wwjjmm@sxu.edu.cn}
\affiliation{%
{${}^{1}$State Key Laboratory of Quantum Optics and Quantum Optics Devices, and Institute of Opto-Electronics, Shanxi University, Tai Yuan 030006, Shanxi Province, People's Republic of China}\\{${}^{2}$Collaborative Innovation Center of Extreme Optics of the Ministry of Education and Shanxi Province, Shanxi University, Tai Yuan 030006, Shanxi Province, People's Republic of China}
}%

\begin{abstract}
We demonstrated stochastic switching in a bistable system implemented with Rydberg atomic ensemble. The transition between the two states of the bistable system is driven by intensity noise of the laser beams. Rydberg atomic ensemble accumulates energy in an equilibrium situation and brings the nonlinear system across the threshold, where stochastic switching occurs between the two states.  Measurement of Rydberg state's population by means of the ladder-type electromagnetically-induced transparency approaching relevant Rydberg state allows us to investigate the nonlinear behavior in Rydberg atomic ensemble based bistable system experimentally. 
\end{abstract}


\maketitle


\section{Introduction}
System resonance is a fundamental phenomenon pervading both nature and society. It reveals the response of a system to the store and transfer of energy from an external forcing source to an internal mode, where the forcing source includes the driving signal and stochastic noise $^{1)}$. Quantum particles have been proposed as powerful constituents that form nonlinear systems. Recent developments in techniques have provided scalable approaches for studying the interplay of pure quantum mechanical systems and their couplings to reservoirs. These techniques can be applied equally to quantum information processing and quantum sensing $^{2)}$.

The phenomenon of stochastic switching in a bistable system has been observed in solid-state crystals, ion systems, and double quantum dot systems $^{3-10)}$. For atomic systems, it has been theoretically predicted that intrinsic interactions lead to stochastic resonances as well. For a neutral atom, the properties of the interaction depend on the quantum state. The interaction of the atoms in their ground state is dominated by 1/$R^6$ van der Waals forces at short range. The excitation of the neutral atom to a high-lying Rydberg state results in strong dipole -dipole and Van der Waals interaction, where long-range cooperative interaction is a promising candidate for implementing a bistable system. Bistability is critical for producing stochastic resonance, where quantum stochastic resonance can be driven by spin noise or quantum fluctuation. Based on nonlinear systems with a cooperative Rydberg interaction, it may be possible to implement atom-based sensing. Electric sensitivity of Rydberg atoms is a promising candidate to achieve weak signal detection. The detection scheme is based on the sensitivity to the initial conditions of either stochastic or chaotic systems, in which the states of the system change under very small perturbations. This is different from the traditional methods.

In our study, we have demonstrated stochastic switching in Rydberg atomic ensemble. The double-state potential system was formed under a cooperative interaction of Rydberg atoms. In this bistable system, population transfer occurs between low and high population of Rydberg states.

\section{Experimental setup}

The experimental apparatus is shown in Fig.1. An 852nm external cavity diode laser (ECDL) is employed as a probe laser, with a typical linewidth of $\sim$MHz. The optical power of 1018 nm ECDL laser is amplified to 5 W by fiber amplifier, and the output beam is frequency-doubled in a PPLN crystal in order to produce a 509 nm laser. The two beams are then overlapped in the cesium atom cell with a counterpropagating configuration. The atomic cell is tens of millimeters in size to match the Rayleigh length of the focused beams: $\sim$250 $\mu$m waist for the 509 nm laser and $\sim$170 $\mu$m waist for the 852 nm laser. The wavelength 852 nm is stabilized to the $6S_{1/2}(F=4)\rightarrow 6P_{3/2}(F'=5)$ hyperfine transitions line via saturated absorption spectroscopy. The wavelength of the probe laser and coupling laser are measured by a wavelength meter (HighFinesse WS-7, wavelength deviation sensitivity is about 2 MHz). The wavelength meter is calibrated by cesium atom hyperfine transition line. The beam of wavelength 509 nm is stabilized through EIT spectroscopy. The two laser servo systems include a kilohertz bandwidth piezoelectric transducer (PZT) and a gigahertz bandwidth bias-Tee current port. The entire feedback bandwidth of the frequency stabilization loop is over 100 kHz. The entire system adopts ultrastable mirror mounts to suppress mechanical noise. The model of the RF function generator (E8257D, Agilent Technologies) is referred to a rubidium clock (FS527, Stanford Research Systems), which is used to drive the electro-optic modulation (EOM). Based on a radio-frequency modulation technique, we have measured the velocity dependence of the hyperfine splitting of intermediate states and the Doppler-free splitting of Rydberg states in a room-temperature vapor cell, as shown in Fig. 2. Using RF modulation technique to measure the relative energy shift requires only an interval of the spectrum, which is accurately distinguishable.

\begin{figure}[htbp]
\centerline{
\includegraphics[width=90mm]{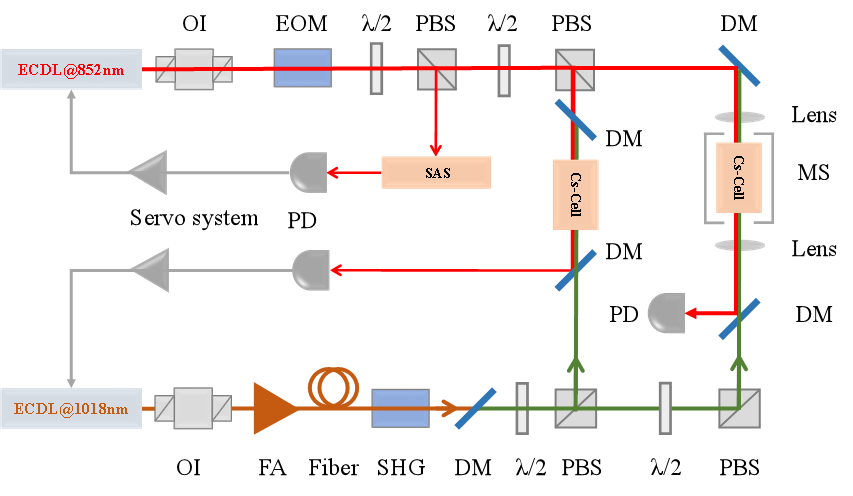}
}
\caption{Schematic of the experimental apparatus. An external cavity diode laser (ECDL) with a wavelength of 852 nm was used as a probe laser. The power of the 1018 nm laser from the ECDL was amplified to 5 W by the fiber amplifier, and the frequency of the output beam was doubled in a periodically poled lithium niobate crystal (PPLN) to produce a 509 nm laser (SHG). Then, the 852 nm and 509 nm laser beams overlapped in the cell (Cs-cell) in a counterpropagating configuration. OI, optical isolator; EOM, electro-optic modulation; $\lambda$/2, half-wave plate; PBS, polarization beam splitter cube; DM, dichroic mirror; MS, Magnetic shielding; FA, fiber amplifier; SHG, second-harmonic generation; MF, Magnetic shield; PD, photodiode detector.}
\label{f1}
\end{figure}

\section{Cooperative Rydberg Interactions}

The nonlinear system produced by neutral atoms requires the preparation of a Rydberg ensemble in special quantum states. Fig. 2(a) shows cascade-type EIT of the Cs atom. The 852 nm probe laser and 509 nm coupling laser are nearly resonant. The EIT transmission signal is monitored by scanning the frequency of the coupling laser while locking the frequency of the probe laser. The strong coupling laser is scanned across the upper transition. The EIT peaks appear when the probe laser and coupling laser are two-photon resonant. The hyperfine splitting of 6$P_{3/2}$ intermediate states can be observed due to the existence of different velocity classes of atoms [Fig. 2(b)]. For atoms with velocity $\vartheta$ moving in the same propagating direction as the probe field, the detuning of the probe laser is $\Delta_p= -\omega_p\cdot \vartheta/c$ and that of the coupling laser is $\Delta_c=\Delta_p\cdot\omega_c/\omega_p$. When the two-photon condition is satisfied, considering Doppler mismatch, the hyperfine splitting of 6$P_{3/2}$ states scale as $\Delta_{Two}=-(1-\omega_c/\omega_p)\cdot\Delta_p$.

\begin{figure}[htbp]
\setlength{\belowcaptionskip}{-0.5cm}
\centerline{
\includegraphics[width=90mm]{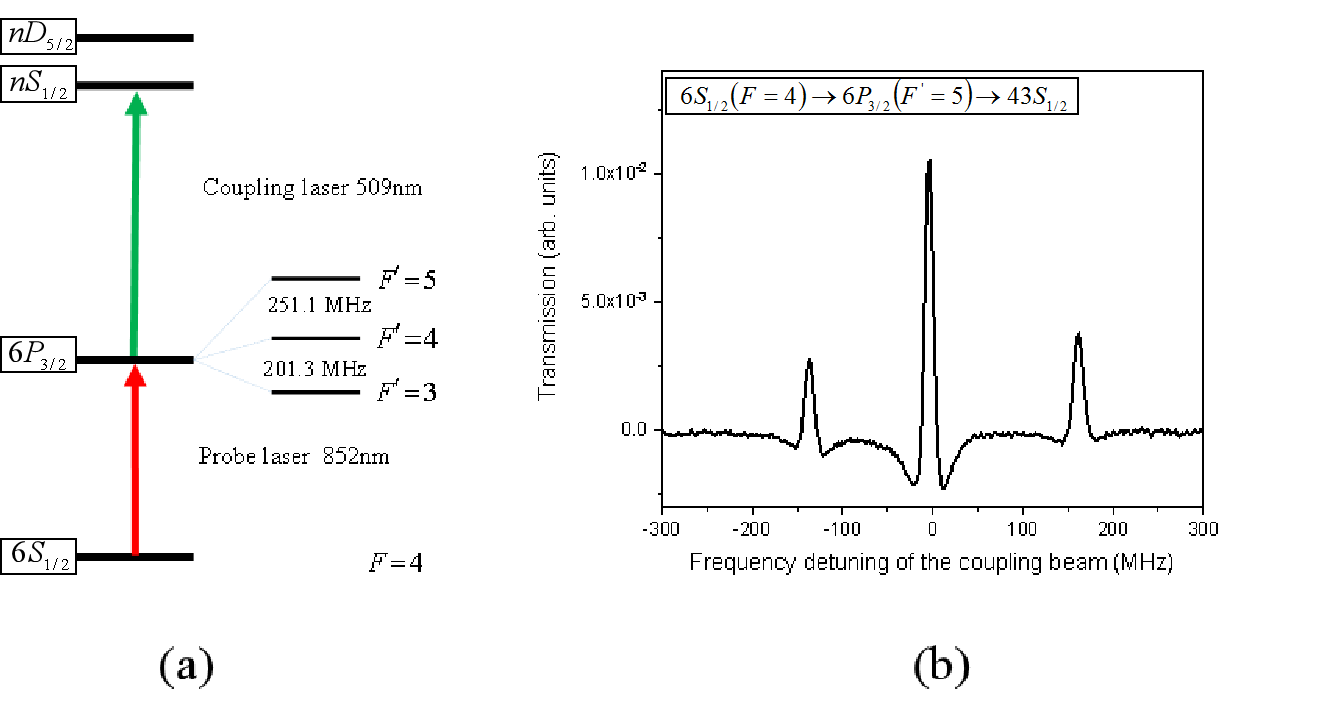}
}
\caption{(a) Energy level schematic of the cascade-type EIT of the Cs atom. The 852 nm probe laser is resonant with the $6S_{1/2}(F=4)\rightarrow 6P_{3/2}(F'=5)$ transition. The 509 nm coupling laser is resonant with the 6P state and the nS state. (b) All hyperfine transition EITs of intermediate states are observed because of velocity-selective effects in the room-temperature atomic cell.}
\label{f2}
\end{figure}

Under the weak probe field regime, the matrix element for the population can be calculated by solving the steady-state optical Bloch equations. The profile of the transmission signal amounts to a convolution of the Holtsmark probability distribution with the EIT line shape: $L(\Delta)=\int{_0^\infty} P(E)\cdot L_{EIT}\cdot dE$$^{10,11)}$. The evolution of the cooperative ensemble can now be described by the modified two-level optical Bloch equations $^{12)}$. Following the Mean-field theory presented by reference $^{13)}$, we simulate the Rydberg population as a function of the coupling laser frequency detuning, and the matrix expressions are given by

\begin{equation*}
{\dot{\rho}_{11}}=\frac{i\Omega}{2}({\rho}_{12}-{\rho}_{21})+\Gamma{\rho}_{22},
\end{equation*}
\begin{equation*}
{\dot{\rho}_{12}}=\frac{i\Omega}{2}({\rho}_{11}-{\rho}_{22})+(i\Delta_{eff}-\Gamma/2){\rho}_{12},
\end{equation*}
\begin{equation*}
{\dot{\rho}_{21}}=\frac{i\Omega}{2}({\rho}_{22}-{\rho}_{11})+(i\Delta_{eff}-\Gamma/2){\rho}_{21},
\end{equation*}
\begin{equation*}
{\dot{\rho}_{22}}=\frac{i\Omega}{2}({\rho}_{21}-{\rho}_{12})-\Gamma{\rho}_{22},
\end{equation*}

The diagonal matrix elements $\rho_{ii}$ represent the population of states and the off-diagonal matrix elements $\rho_{ij}$ represent the coherence between the states. $\Gamma$ is the decay rate from the Rydberg state. The atoms in ground state is pumped to Rydberg state by probe laser and coupling laser. Taking into account the two-photon transition, here the effective Rabi frequency $\Omega$ and effective detuning $\Delta_{eff}$ have been considered. This equation can be solved numerically to obtain the steady-state Rydberg population. The expressions are given by

\begin{equation*}
(1-2{\rho}_{22})^3(\frac{V^2}{4})+(1-2{\rho}_{22})^2(\Delta V-\frac{3V^2}{4}) 
\end{equation*}

\begin{equation*}
+(1-2{\rho}_{22})(\frac{3V^2}{4}-2\Delta V+{\Delta}^2+\frac{{\Gamma}^2}{4}+\frac{{\Omega}^2}{2})
\end{equation*}

\begin{equation*}
+(\Delta V-{\Delta}^2-\frac{V^2}{4}-\frac{{\Gamma}^2}{4})=0
\end{equation*}

When there is weak interaction in the Rydberg ensemble, the transmission signal is described by a Lorentzian profile [gray dotted line in Fig. 3(a)]. When there are strong interactions, the system response is governed by quantum state-dependent nonlinear dynamics. Here, the Rabi frequency of the coupling laser and the probe laser are both 11.5 MHz. The interaction energy is $\pm$49 MHz. At critical frequency of laser, there is a sharp switching in the Rydberg population [green dash lines in Fig.3(a)]. The atomic density-dependent cooperative interaction depends on the power and detuning of driving lasers. We use the EIT to measure the population-dependent effect. The transmission spectroscopy of $6S_{1/2}(F=4)\rightarrow 6P_{3/2}(F'=5)\rightarrow 50D_{5/2}$ present non-symmetric profiles as the Rabi frequency of probe laser being increased from $\sim$3.4 MHz to $\sim$12.5 MHz, as shown in Fig. 3(b). Cooperative interaction results in the nonlinear transition effect.  Here, the observed peaks shift may be also due to the ion-dependent energy shift and Cooperative interaction shift. The frequency of the probe laser is optimized by blue detuning $\sim$200 MHz relative to the $6S_{1/2}(F=4)\rightarrow 6P_{3/2}(F'=5)$transition frequency. Taking into account the Doppler mismatch, the corresponding frequency detuning of coupling laser is  $\Delta_{Two}=-(1-\omega_c/\omega_p)\cdot\Delta_p\simeq-135MHz$. Note that the nonlinear collective interaction is sensitive to the ground state atomic density. The optimized cell temperature of the atomic gas is 35$^{\circ}$C, which corresponds to an atomic density of $\sim1.3\times10^{11}cm^{-3}$.

\begin{figure}[htbp]
\setlength{\belowcaptionskip}{-0.5cm}
\centerline{
\includegraphics[width=95mm]{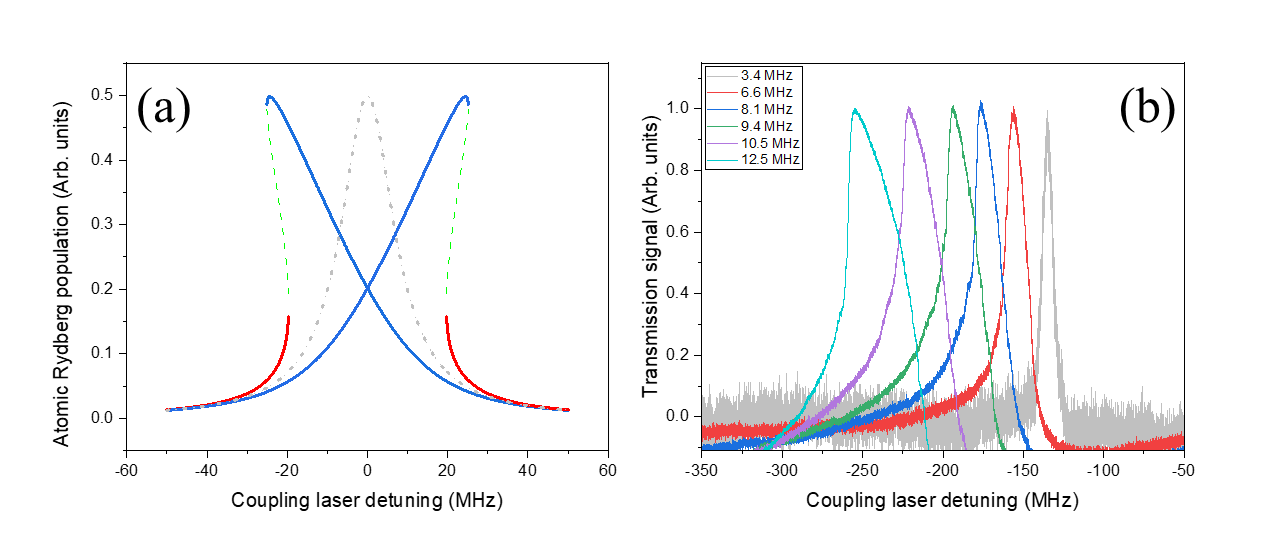}
}
\caption{(a) The model of the cooperative interaction in a Rydberg ensemble as a function of coupling laser frequency detuning. The gray dotted line is a Lorentzian profile shape. The red and blue lines plot the low and high Rydberg populations, respectively. The sign of the interaction frequency shift relative to the unperturbed resonance depends on the angular momentum states. (b) The transmission spectroscopy of   $6S_{1/2}(F=4)\rightarrow 6P_{3/2}(F'=5)\rightarrow 50D_{5/2}$ present non-symmetric profiles, the Rabi frequency of probe laser are increased from $\sim$3.4 MHz to $\sim$12.5 MHz.}
\label{f3}
\end{figure}

Fig. 4 shows the dependence of the nonlinear effects on coupling laser intensity and frequency. The Rabi frequency of the coupling laser beams are increased from $\sim$5.1 MHz to $\sim$12.7 MHz. When the power of coupling laser is weak, the profile of transmission signal are similar to a Lorentzian function or a Voigt function. In case of strong laser power, at critical frequency, there is a sharp switching in the transmission signal. The switching of nonlinear effect is caused by cooperative interactions in the Rydberg atomic ensemble. These interactions include the dipole–dipole interaction between Rydberg atoms $^{12-14)}$, the charge-induced interaction between the Rydberg atoms and the charge produced by the spontaneous ionization of the Rydberg atoms $^{10,15-18)}$. The hysteresis phenomenon of Rydberg population depends on the laser scanning direction. The frequency of the probe laser is optimized by blue detuning of $\sim$200 MHz relative to the $6S_{1/2}(F=4)\rightarrow 6P_{3/2}(F'=5)$transition frequency. The probe laser is blue detuned by $\sim$450 MHz relative to the $6S_{1/2}(F=4)\rightarrow 6P_{3/2}(F'=4)$transition frequency, where the far-off resonance condition significantly reduces the pumping effects of Rydberg population. This may be the reason why there is no nonlinear effect for $6S_{1/2}(F=4)\rightarrow 6P_{3/2}(F'=5)\rightarrow 50D_{5/2}$transition. In contrast to the power-dependent effect of probe laser, as shown in Fig 3(b), the transmission windows of the EIT signal are not obviously shifted. The peaks shift may arise from Stark shift due to the spontaneous ionization of the Rydberg atoms $^{10,18-20)}$. In addition, strong driving from the probe laser inhibits the ensemble occupation of the Rydberg state, which effectively suppresses avalanche ionization, resulting in a shift in the EIT window with probe laser power $^{21)}$.

\begin{figure}[htbp]
\setlength{\belowcaptionskip}{-0.5cm}
\centerline{
\includegraphics[width=90mm]{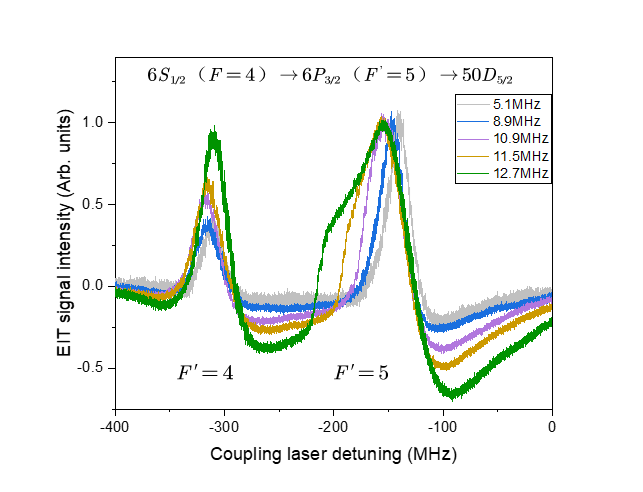}
}
\caption{Time series plots for two 6P hyperfine states with various coupling laser intensity. Distinctive nonlinear characteristics from EIT spectroscopy are observed for the transition  $6S_{1/2}(F=4)\rightarrow 6P_{3/2}(F'=5)\rightarrow 50D_{5/2}$. The probe laser is optimized by blue detuning of $\sim$200 MHz relative to the $6S_{1/2}(F=4)\rightarrow 6P_{3/2}(F'=5)$ transition frequency. The Rabi frequency of the probe laser is $\sim$ 4 MHz. }
\label{f4}
\end{figure}
\section{Stochastic switching in a bistable system }
Stochastic switching in a bistable system is similar to the threshold-crossing excitable system. The system accumulates energy in the resting state. Perturbations above a certain threshold may induce large excursions triggering transitions. When noise is injected into a bistable system, the collective interaction assists the intrinsic oscillator in eliciting an efficient response by overcoming the potential barrier. Then, the resting condition is transformed into a firing condition. The bistable system can be described by the universal scaling theory of the FitzHugh–Nagumo model $^{22)}$, which is a simplified version of the Hodgkin–Huxley model $^{23,24)}$. Similarly, a Rydberg ensemble with long-range cooperative can also perform the nonlinear system. In case of low atomic density, the Rydberg excitations are independent, and the mutual interactions do not affect the system dynamics. As the atomic density increases, the interaction of the nearest neighbors strongly prevents the generation of new Rydberg atoms. When the number of interacting Rydberg atoms is above a critical number, the cooperative shift exceeds the width of the electron shelving resonance where the system state does not change with the laser frequency detuning in a finite range. This results in low and high Rydberg population $^{12,13)}$. The hysteresis phenomenon of Rydberg population depends on the laser scanning direction. The low and high Rydberg populations form the two stable states of system. When the driving energy is small, there is no cross motion. When the cumulative energy of the system is above the potential barrier, switching between the two states occurs.

\begin{figure}[htbp]
\setlength{\belowcaptionskip}{-0.5cm}
\centerline{
\includegraphics[width=100mm]{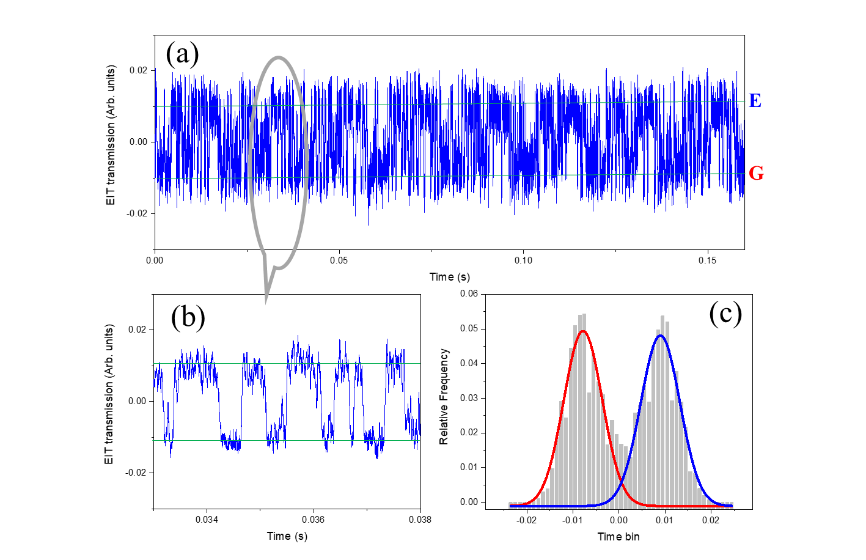}
}
\caption{Typical stochastic resonance spectroscopy: (a) Noise-induced switching traces in the nonlinear system produced by the cooperative interaction of the Rydberg ensemble. G and E indicate the transmission signals for the populations of the low and high Rydberg states, respectively; (b) More detailed results, the step-like signal with detail is almost stochastic switching; (c) Histogram of the photon counting data—the discretely separated count peaks (red and green fitted lines) represent two equilibrium points in the Rydberg ensemble system. Both peaks are fitted by a Gaussian function.}
\label{f5}
\end{figure}

The nonlinear interaction of the Rydberg atoms creates a two-state system. The barrier heights of the two-state can be controlled by tuning the laser parameters. The laser beams with white Gaussian noise provide the perturbations or driving force. When the lasers derive the atomic Rydberg ensemble, the noise of laser acts on the bistable system, the cumulative energy of the system in the equilibrium state is above the barrier height, and a transition occurs between the two states of the bistable system. Fig. 5(a) shows the experimental implementation of the stochastic switching for Rydberg atomic ensemble in the 50$D_{5/2}$ state, as detailed in Fig. 5(b). The spectroscopy of a time series plot is obtained by tuning the coupling laser frequency. The tolerance frequency range of stochastic switching is less than 10 MHz. The Rabi frequencies of the probe laser and the coupling laser are $\sim$11.46 MHz and $\sim$11.48 MHz, respectively. In the experiments, the frequency of the probe laser is blue detuned by about 200 MHz relative to the $6S_{1/2}(F=4)\rightarrow 6P_{3/2}(F'=5)$transition frequency

 Almost all of the stochastic switching characteristics originate from the transition between the populations of the low and high Rydberg states. The average duration of these states are $\sim$ 174 $\mu$s and $\sim$ 178 $\mu$s, respectively. Typical rise and fall times are $\sim$ 66 $\mu$s and $\sim$ 83 $\mu$s, respectively. Fig. 5(c) shows the frequency counts of the transmission signal. The relative occurrences number of bistability states are  49.5$\%$ and 50.5$\%$. Clearly well-separated peaks demonstrate that there are two equilibrium points in the Rydberg ensemble system. The wide range of state distributions is due to system fluctuations and the conversion of the laser phase noise to amplitude noise in the EIT system. The transition characteristics are analogous to the excitation of the two-level atom system. The bistable system is subjected to a certain amount of internal or external random perturbations. State coherence is characterized by autocorrelations, where coherence times of $\sim$ 128 $\mu$s are typical. The aperiodicity character shows that the transitions are driven by a stochastic distribution noise. 

\section{Summary and outlook }

We explored the stochastic switching of a bistable system in the Rydberg ensemble. The cooperative interaction of the Rydberg atoms produces a bistable double-state system. The collective state of the Rydberg atoms accumulates energy in an equilibrium situation. External noise occasionally gives the system a kick that is large enough to cross the barrier of the double-state. An indicator of stochastic resonance is that the flow of information through a system is maximized when the input noise intensity matches the system response, which is one of the fundamental laws in physics, engineering, and biology. A quantum nonlinear system will be useful for implementing resonance sensing and precision measurements. Note here the cooperative interactions of the Rydberg ensemble are used to establish a bistable system, but the driving force of the stochastic switching is not quantum noise. Nevertheless, stochastic resonance driven by quantum noise has not been realized. The proposal of Rydberg ensemble bistability does realize quantum stochastic resonance with spin noise.

\vspace{0.2 in}
\noindent\textbf{Funding.} This work was supported by the National Natural Science Foundation of China (Grant Nos. 61875111, 11774210, 61905133, and 11974226), the National Key Research and Development Program of China (2017YFA0304502), the Shanxi Provincial 1331 Project for the Key Subject Construction, and the Scientific and Technological Innovation Programs of Higher Education Institutions in Shanxi (2017101).

\end{document}